%% file: neurips_2026.tex
\documentclass{article}

 \usepackage{amsmath}
\usepackage[preprint]{neurips_2026}

\usepackage[utf8]{inputenc} 
\usepackage[T1]{fontenc}    
\usepackage{hyperref}       
\usepackage{url}            
\usepackage{booktabs}       
\usepackage{amsfonts}       
\usepackage{nicefrac}       
\usepackage{microtype}      
\usepackage{xcolor}         
\usepackage{multirow}
\usepackage{enumitem}
\usepackage{graphicx}
\usepackage{subcaption}
\usepackage[table]{xcolor}
\usepackage{wrapfig}
\usepackage{tcolorbox}

\title{CompRank: Efficient LLM Reranking via Token-Level Compression and Decoding-Free Scoring}

\author{%
 \textbf{Xuan Lu\textsuperscript{1,2,3}},\;
 \textbf{Haohang Huang\textsuperscript{3}},\;
 \textbf{Yingqi Fan\textsuperscript{3}},\;
 \textbf{Junlong Tong\textsuperscript{1,2,3}},\;
 \textbf{Yuxuan Zhang\textsuperscript{4}},\;\\
 \textbf{Ping Nie\textsuperscript{5}},\;
 \textbf{Rui Meng}\thanks{Working at Google Cloud AI Research. $\ddagger$ Corresponding author.},\;
 \textbf{Xiaoyu Shen\textsuperscript{2,3}$^{\ddagger}$}\\[4pt]
 \textsuperscript{1}Shanghai Jiao Tong University\\
 \textsuperscript{2}Ningbo Key Laboratory of Spatial Intelligence and Digital Derivative\\
 \textsuperscript{3}Institute of Digital Twin, Eastern Institute of Technology, Ningbo\\
\textsuperscript{4}University of British Columbia
\textsuperscript{5}University of Waterloo\\
 \texttt{lux1997@sjtu.edu.cn\quad xyshen@eitech.edu.cn}\\[3pt]
}

\begin{document}

\maketitle

\begin{abstract}
Large language model (LLM) rerankers have become an important component of modern retrieval and retrieval-augmented generation pipelines, but their high computational cost limits their applicability to long candidate lists. In this paper, we propose \textbf{CompRank}, a token-efficient reranking framework that reduces redundant computation by aligning reranker design with the sparsity of ranking signals. CompRank decouples document representations from candidate order and query context, enabling reusable document-side states; applies segment-wise token compression to reduce query--document interaction cost; and introduces a CopyNet-style objective that directly aligns attention-based document scoring with training supervision. Experiments on seven BEIR datasets show that CompRank achieves strong reranking performance while retaining only 10.2\% of document tokens, reaching an average NDCG@10 of 39.2 compared with 39.7 under full-token attention. Further scaling experiments on TREC-COVID show that CompRank remains stable when evaluated on candidate lists of up to 500 documents after training on 30-document lists, while achieving $4.9\times$--$9.5\times$ end-to-end speedup over generation-based listwise reranking and approximately $1.3\times$ speedup over the full-token CompRank variant. These results suggest that token-level compression and decoding-free attention scoring provide an effective path toward scalable LLM-based reranking.
\end{abstract}

\input{sec/01_Intro}
\input{sec/02_Related_Work}

\input{sec/03_Methods}
\input{sec/04_Experiments}

\input{sec/05_Analysis}
\input{sec/06_Conclusion}








{\small
    \bibliographystyle{plain}
    \bibliography{neurips_2026}
}

\input{sec/X_appendix}


\end{document}

%% file: sec/01_Intro.tex
\section{Introduction}

Reranking has become an important component of modern Information Retrieval (IR) and Retrieval-Augmented Generation (RAG) pipelines~\cite{rag,lumulticonir}. By applying fine-grained relevance modeling to candidates retrieved by first-stage models, rerankers, especially those based on Large Language Models (LLMs), can improve retrieval precision and support downstream agentic decision-making~\cite{lu2026tools,huang2026mmeb}. However, these gains often come with substantial computational cost. In long-list scenarios, such as reranking the top-100 retrieved documents, LLM-based rerankers must process long input contexts and perform expensive attention computation, making latency and memory consumption major bottlenecks.
Recent studies have explored several ways to improve the efficiency of LLM-based reranking. Direct-Rank~\cite{lu2026rethinking} shows that ranking does not necessarily require lengthy Chain-of-Thought reasoning, and that direct ranking sequence modeling can achieve strong performance. To reduce decoding overhead, ICR~\cite{chen2025attention} and QRHead~\cite{QRHead} infer relevance scores from internal attention patterns, avoiding expensive autoregressive generation. BlockRank~\cite{blockrank} and LongRanker~\cite{zhou2026longranker} further observe redundancy in long-context reranking and exploit structured attention to reduce unnecessary interactions. These methods improve efficiency from different perspectives, but many of them still process dense token-level representations, leaving substantial redundancy in document-side computation and query--document interaction.

In this work, we revisit reranking from the perspective of token-level sparsity and representation reuse. Our motivation is that relevance estimation often depends on a relatively small subset of informative document tokens, while the remaining tokens contribute limited signal for ranking. At the same time, candidate documents are frequently recomputed in different contexts, even when their representations need not depend on the current query or candidate ordering. This issue is particularly relevant in agentic search and RAG systems, where agents may revisit the same documents, maintain long interaction histories, and repeatedly invoke retrieval tools across multi-step reasoning trajectories. In such settings, reusable document-side KV states provide a natural way to amortize repeated document encoding and reduce online computation. This creates a mismatch between dense computation and sparse ranking signals. We hypothesize that reranking can be made more efficient by decoupling document representations and reducing the number of document tokens involved in query-side scoring.
To this end, we propose \textbf{CompRank}, a token-efficient reranking framework for long candidate lists. CompRank introduces a document representation decoupling mechanism that reduces dependence on candidate order and query context, making reusable document representations possible. It then applies structured token compression to reduce the number of document tokens exposed to query--document attention. Finally, CompRank uses a CopyNet-style objective to align training with attention-based inference: the same attention-derived document scores are optimized during training and directly used for decoding-free reranking at inference time.

This design provides both efficiency and robustness benefits. Since CompRank avoids autoregressive decoding of document identifiers, it reduces decoding latency and avoids common generation failures such as duplicated or out-of-range document IDs. Since document tokens are compressed before query-side scoring, it further reduces the effective attention cost. Moreover, because ranking is performed over input candidates rather than generated output sequences, CompRank can better extrapolate from shorter training lists to longer inference-time candidate lists.
We evaluate CompRank on seven BEIR datasets. With Segment-10 compression, CompRank retains only 10.2\% of document tokens while achieving an average NDCG@10 of 39.2, close to the full-token setting of 39.7. In scaling experiments on TREC-COVID, CompRank generalizes from 30-document training lists to 500-document inference lists and achieves $4.9\times$--$9.5\times$ end-to-end speedup over a generation-based listwise reranker, as well as approximately $1.3\times$ speedup over the full-token CompRank variant.

Our contributions are summarized as follows:
\begin{itemize}
    \item We propose CompRank, a token-efficient reranking framework that decouples document representations and reduces redundant computation through structured token compression.
    \item We align attention-based inference with CopyNet-style supervision, directly optimizing the same attention-derived document scores used for decoding-free reranking.
    \item We show that CompRank remains more stable than decode-ID reranking in our TREC-COVID scaling experiment when evaluated on longer candidate lists.
    \item We provide complexity and latency analyses, showing reduced attention cost and practical end-to-end speedups over generation-based and full-token reranking baselines.
\end{itemize}

%% file: sec/02_Related_Work.tex





\section{Related Work}

\paragraph{LLMs for Reranking}

Recent advances have established large language models (LLMs) as a dominant paradigm for reranking~\citep{yin2026queriesdecomposedstageawarestudy,lu2026beyond,fan2026minirerankerefficientmultimodalreranking}. Existing approaches can be broadly categorized into pointwise~\citep{lu2026rethinking}, pairwise~\citep{qin-etal-2024-large}, setwise~\citep{zhuang2024setwise}, and listwise methods~\citep{pradeep2023rankvicunazeroshotlistwisedocument}.
Pointwise methods evaluate each query--document pair independently by generating relevance scores, with representative models including MonoBERT~\citep{monobert}, MonoT5~\citep{monot5}, and RankLLaMA~\citep{rankllama}. While effective, this paradigm incurs substantial redundant computation on large candidate sets. Pairwise and setwise methods model relative preferences among documents~\citep{zhuang2024setwise,qin-etal-2024-large}, but introduce additional overhead due to repeated comparisons.
Listwise and in-context ranking approaches jointly process multiple candidates within a single context, enabling richer inter-document interactions. Early work demonstrated strong zero-shot performance using GPT-based rerankers~\citep{sun2023rankgpt}, followed by adaptations to open-source LLMs such as Vicuna, Zephyr, and Mistral~\citep{pradeep2023rankvicunazeroshotlistwisedocument,pradeep2023rankzephyreffectiverobustzeroshot}, as well as structured frameworks like RankLLM~\citep{sharifymoghaddam2025rankllm}. To scale to long candidate lists, these methods often employ sliding windows or hierarchical strategies~\citep{zhou2026longranker}, which reduce context length but introduce redundant computation across windows.

\paragraph{Efficient Rerankers}

Improving the efficiency of long-context reranking has attracted increasing attention. One line of work reduces decoding cost. Attention-based rerankers infer relevance directly from internal attention patterns without explicit generation~\citep{chen2025attention,QRHead,wang2026headrankdecodingfreepassagereranking,tran2026contrastiveretrievalheadsimprove}, while FIRST reduces listwise reranking to single-token decoding~\citep{first}. These methods avoid or shorten autoregressive ranking outputs, thereby reducing inference overhead.
Another line of work reduces input or attention computation. BlockRank~\citep{blockrank} shows that inter-document attention is largely redundant and proposes block-wise attention for scalable in-context ranking. LongRanker~\citep{zhou2026longranker} studies efficient one-pass reranking under long-context settings. General long-context models also explore sparse attention patterns such as sliding-window or global-local attention~\citep{beltagy2020longformer,zaheer2020bigbird}, and recent KV-level compression methods reduce the number of tokens participating in attention~\citep{deepseekai2024deepseekv32,deepseekai2026deepseekv4}.
A complementary direction compresses or reuses document representations. HyperRAG~\citep{hyperRAG}, CompressedDoc~\citep{dejean2025rerankingcompresseddocumentrepresentation}, and BlockDoc~\citep{li2026efficientlongdocumentrerankingblocklevel} precompute or compress document representations for reuse. More closely related, Compress-then-Rank applies ranking-aware passage compression before listwise reranking~\citep{zhi2026compress}, and PE-Rank uses passage embeddings as compact surrogates for efficient listwise reranking~\citep{liu2025leveraging}. In contrast, CompRank preserves token-level document representations, compresses the subset of document tokens exposed to attention-based scoring, and performs decoding-free reranking by directly sorting attention-derived document scores.

%% file: sec/03_Methods.tex
\begin{figure}
    \centering
    \includegraphics[width=0.99\linewidth]{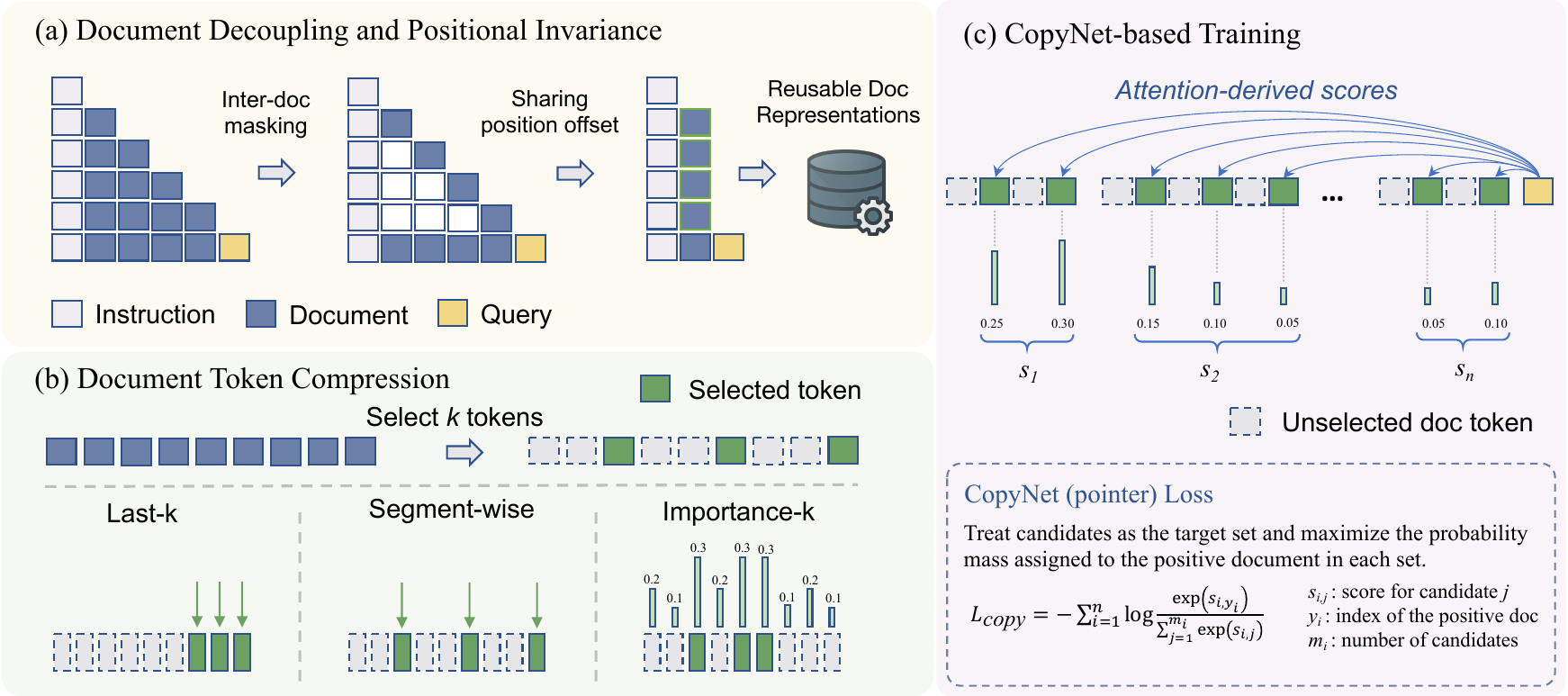}
\caption{Overview of CompRank. 
(a) Document blocks are decoupled by inter-document masking and shared positional offsets, enabling reusable document representations.
(b) Token-level compression reduces the number of document tokens exposed to query-side attention.
(c) CopyNet-style supervision directly optimizes attention-derived document scores for decoding-free reranking.}
    \label{fig:comprank_framework}
\end{figure}

\section{CompRank}
\label{sec:method}

In this section, we present \textbf{CompRank}, an efficient framework for long-list reranking. Our design is motivated by three objectives: (i) avoiding the latency of autoregressive decoding through attention-based scoring; (ii) reducing sensitivity to candidate order and enabling document-side pre-computation through representation decoupling; and (iii) reducing query--document interaction cost through token-level compression.

\subsection{Problem Formulation: Scoring via Attention Mass}

Recent work~\cite{chen2025attention,QRHead} shows that reranking can be performed directly from the contextual encoding stage of LLMs, without relying on autoregressive generation. In particular, in-context reranking (ICR) is based on the observation that, when processing a query together with candidate documents, LLMs often allocate more attention mass to tokens belonging to relevant documents. This suggests that relevance can be estimated by aggregating attention weights from query-side tokens to document-side tokens.

Formally, given a query $q$ and a candidate document set $\mathcal{D}=\{d_1,d_2,\ldots,d_N\}$, the goal is to assign a relevance score to each document and produce a ranked list. Building on ICR, CompRank formulates reranking as a \textit{discriminative attention aggregation} problem, where relevance scores are derived directly from attention distributions without decoding document identifiers.

Let $B_q$ denote the query or decision block, and let $T_i$ be the token sequence of document $d_i$. After optional token compression (Sec.~\ref{sec:compression}), the visible token subset is denoted as $T_i' \subseteq T_i$. For a scoring layer $\ell^*$ and attention head $h$, the masked attention logit from a query-side decision token $u \in B_q$ to a document token $t \in T_i'$ is:
\[
a^{(\ell^*,h)}_{u,t} =
\frac{
Q^{(\ell^*,h)}_u
\left(K^{(\ell^*,h)}_{t}\right)^\top
}{
\sqrt{d_h}
}
+
m_{u,t},
\]
where $Q$ and $K$ are query and key representations, $d_h$ is the head dimension, and $m_{u,t}$ denotes the attention mask. We normalize attention over all retained document tokens:
\[
p^{(h)}_{u,t} =
\frac{
\exp(a^{(\ell^*,h)}_{u,t})
}{
\sum_{j=1}^{N}
\sum_{t'\in T_j'}
\exp(a^{(\ell^*,h)}_{u,t'})
}.
\]
The document-level score for $d_i$ is obtained by aggregating attention mass over its retained tokens:
\[
s_i =
\operatorname{Agg}_{u\in U}
\left[
\frac{1}{H}
\sum_{h=1}^{H}
\sum_{t\in T_i'}
p^{(h)}_{u,t}
\right],
\]
where $U \subseteq B_q$ is the set of decision tokens, $H$ is the number of heads, and $\operatorname{Agg}(\cdot)$ is an aggregation operator such as mean or max. The final ranking $\pi$ is obtained by sorting documents in descending order of $s_i$. Unless otherwise specified, the scoring layer, decision-token set, and aggregation operator are fixed across all datasets; implementation details are provided in Appendix~\ref{app:implementation_details}.

\subsection{Document Representation Decoupling and Positional Invariance}

A key challenge in in-context reranking is the sensitivity of LLMs to the order of candidate documents. Since candidate lists are dynamically constructed and may vary across queries, reducing order sensitivity is important for robust reranking. CompRank addresses this by organizing the input as independent document blocks followed by a query or decision block:
\[
X = [D_1; D_2; \dots; D_N; B_q].
\]

To reduce dependence on candidate ordering, we use a shared positional bias across document blocks. Specifically, for a token $x_{i,t}$ at local position $t$ in document block $D_i$, we assign:
\[
\operatorname{pos}(x_{i,t}) = t,
\]
independent of the document index $i$. For the query block $B_q$, we apply a constant offset $C$:
\[
\operatorname{pos}(B_q,t)=C+t.
\]
This removes document-index-dependent positional offsets and reduces one major source of order sensitivity in document representations.

Following the block-structured attention design in BlockRank~\cite{blockrank}, we further impose a structured attention mask to decouple candidate documents. Tokens within each document block perform causal self-attention, while attention across different document blocks is prohibited. The query block $B_q$ attends to all document blocks, maintaining a global view over the candidate set for relevance estimation.

Under this design, document representations are computed independently of other candidates and do not depend on the query block. This makes document-side pre-encoding possible in principle:
\[
K_i,V_i=f_\theta(D_i),
\]
where the resulting key-value states can be reused across different queries or candidate permutations. This decoupling reduces repeated document-side computation and provides a path toward more efficient query-time reranking with reusable document states.

\subsection{Token-Level Compression}
\label{sec:compression}

Although document representation decoupling removes document--document interactions, query-side attention still scales with the total number of document tokens. CompRank therefore applies token-level compression to the document-side KV states by selecting a subset $T_i' \subseteq T_i$ of visible document tokens. Importantly, compression is applied only to query-to-document attention in the final block, while document-side representations remain unchanged.

We consider three token selection strategies:
\begin{itemize}
    \item \textbf{Last-$k$ Compression:}
    Retains the final $k$ valid tokens of each document, motivated by the observation that later positions in decoder-only LLMs may aggregate preceding context.

    \item \textbf{Segment-wise Compression:}
    Retains one token every $k$ positions, either start-anchored or end-anchored. This provides uniform coverage across the document while reducing the visible KV size by approximately a factor of $k$.

    \item \textbf{Intrinsic Importance Selection:}
    Uses document-internal attention to select salient tokens. Specifically, the final token of document $d_i$ is used as a summary query to compute token importance:
    \begin{equation}
    \alpha_{i,t}
    =
    \frac{1}{H}
    \sum_{h=1}^{H}
    \frac{
    Q^{(h)}_{i,L_i}
    \left(K^{(h)}_{i,t}\right)^\top
    }{
    \sqrt{d_h}
    }.
    \end{equation}
    We then define $T_i'$ as the union of the top-$k$ tokens according to $\alpha_{i,t}$ and a small set of anchor tokens, such as the last token, to preserve structural information.
\end{itemize}

\subsection{CopyNet-Based Learning Objective}

Standard supervised fine-tuning for LLM reranking often trains the language modeling head to generate document identifiers. This formulation is inefficient for reranking because probability mass is assigned over the full vocabulary, including invalid or irrelevant tokens. It also introduces autoregressive decoding overhead and may generalize poorly when the candidate set size changes.

CompRank instead performs reranking directly using attention-derived document scores. We adopt a \textbf{CopyNet-style objective}~\cite{copynet}, where the model selects from existing input candidates rather than generating unseen labels. Given document scores $\{s_i\}_{i=1}^N$, we define a pointer distribution over candidates:
\[
P_{\mathrm{copy}}(i \mid q,\mathcal{D})
=
\frac{
\exp(s_i/\tau)
}{
\sum_{j=1}^{N}
\exp(s_j/\tau)
},
\]
where $\tau$ is a temperature parameter. For a query with ground-truth document index $y$, the training objective is:
\[
\mathcal{L}_{\mathrm{copy}}
=
-\log P_{\mathrm{copy}}(y \mid q,\mathcal{D}).
\]

This objective directly supervises the same attention-derived scores used at inference time. As a result, training and inference are aligned: the model learns to allocate probability mass over candidate documents, and reranking is performed by sorting these scores without autoregressive decoding. This candidate-level formulation also avoids invalid document identifier generation and supports more stable extrapolation to candidate set sizes beyond those seen during training.

%% file: sec/04_Experiments.tex
\section{Experiments}
\label{sec:experiments}

This section presents the experimental setup, the main results of CompRank, and the comparison of different token compression strategies.

\subsection{Experiment Setup}

We evaluate CompRank on seven datasets from the BEIR benchmark~\cite{thakur2021beir}, covering diverse retrieval tasks and domains: FiQA, SciFact, TREC-COVID, DBPedia, Climate FEVER, NF Corpus, and SciDocs. For comparison, we consider representative retrieval and reranking baselines, including the sparse retriever BM25, generation-based listwise rerankers such as RankGPT~\cite{rankgpt}, and attention-based in-context reranking methods including ICR~\cite{chen2025attention} and Corehead~\cite{tran2026contrastiveretrievalheadsimprove}.
For evaluation, BM25 is used to retrieve the top-100 candidate documents for each query, and all reranking methods reorder the same candidate set. We report NDCG@10 as the primary evaluation metric. 

CompRank uses Mistral-7B as the backbone model and is trained with full-parameter fine-tuning. We directly adopt the training data constructed by BlockRank~\cite{blockrank}, which provides query--document candidate sets with annotated positive instances and is derived from the MS MARCO corpus by sampling 10\% of the original training split. During training, each query is paired with 30 candidate passages retrieved by a first-stage retriever. Unless otherwise specified, we use segment-wise compression with Step-10 as the default compression strategy. We train CompRank with a learning rate of $3.0 \times 10^{-6}$, and all training and evaluation experiments are conducted on 8 NVIDIA H20 GPUs.




\subsection{Main Results}

\paragraph{Comparison of Token Compression Strategies}
We conduct an ablation study to examine how different token compression strategies affect reranking performance. Table~\ref{tab:compression_ablation} reports NDCG@10 on BEIR datasets, together with token retention ratios during inference. The full-token setting achieves the highest average score of 39.7, serving as the upper-bound reference, but requires attending to all document tokens. In contrast, compressed variants substantially reduce the token budget while maintaining competitive performance, indicating considerable token-level redundancy in reranking.
Overall, segment-wise compression provides the best effectiveness--efficiency trade-off. Among all compressed variants, \textbf{Step-10} achieves the highest average score of 39.2, only 0.5 points below full-token attention, while retaining merely \textbf{10.2\%} of document tokens. This suggests that a small uniformly sampled subset can preserve most ranking signals. Increasing the budget to Step-5 (20.0\%) does not yield further gains, while stronger compression with Step-20 and Step-30 gradually degrades performance, showing a clear trade-off between coverage and compression strength.
By contrast, last-$k$ compression exhibits a consistent gap. Although increasing $k$ improves the average score from 36.3 to 38.0, even Last-32 with 14.1\% retained tokens remains below Step-10 with 10.2\%. This indicates that relying only on final tokens introduces positional bias and may discard evidence appearing earlier in the document. Attention-based Top-$k$ selection shows similar limitations: pure Top-$k$ variants obtain 35.8--36.8 on average, and hybrid variants provide limited improvements. These results suggest that document-internal saliency does not reliably align with query-specific relevance.

\begin{table*}[ht]
\centering
\small
\setlength{\tabcolsep}{4pt}
\renewcommand{\arraystretch}{1.15}
\resizebox{\textwidth}{!}{
\begin{tabular}{lccccccccc}
\toprule
\textbf{Strategy} & \textbf{Token Ratio}
& \textbf{Climate} & \textbf{DBPedia} & \textbf{FiQA} & \textbf{NFCorpus} 
& \textbf{SciDocs} & \textbf{SciFact} & \textbf{COVID} & \textbf{Avg.} \\
\midrule
\multicolumn{10}{c}{\textbf{No Compression}} \\
\midrule
\rowcolor[rgb]{  .949,  .949,  .949}
Full Tokens & 100\%
& 16.9 & 35.8 & \textbf{33.1} & \textbf{32.7} & 13.7 & \textbf{67.2} & 78.4 & 39.7 \\
\midrule
\multicolumn{10}{c}{\textbf{Last-$k$ Token Compression}} \\
\midrule
Last-1  & 0.4\% & 16.7 & 34.4 & 29.4 & 30.1 & 12.8 & 52.1 & 78.8 & 36.3 \\
Last-2  & 0.9\% & 17.2 & 34.8 & 30.8 & 29.9 & 13.0 & 53.5 & 78.8 & 36.9 \\
Last-4  & 1.8\% & 17.4 & \textbf{36.1} & 28.6 & 30.4 & 13.4 & 55.4 & 79.7 & 37.3 \\
Last-8  & 3.5\% & 16.2 & 35.4 & 31.5 & 31.3 & 12.9 & 55.8 & 79.9 & 37.6 \\
Last-16 & 7.0\% & 16.0 & 35.2 & 31.5 & 31.4 & 12.9 & 57.4 & 79.6 & 37.7 \\
Last-32 & 14.1\% & 15.6 & 35.3 & 32.5 & 31.0 & 13.4 & 59.5 & 78.9 & 38.0 \\
\midrule
\multicolumn{10}{c}{\textbf{Attention-based Top-$k$ Compression}} \\
\midrule
Top-4  & 1.8\% & 16.8 & 32.5 & 28.0 & 29.1 & 12.8 & 54.7 & 76.9 & 35.8 \\
Top-8  & 3.5\% & 16.6 & 33.2 & 28.8 & 29.7 & 12.9 & 56.0 & 77.4 & 36.3 \\
Top-16 & 7.0\% & 16.2 & 33.6 & 29.6 & 30.1 & 13.0 & 57.6 & 77.7 & 36.8 \\
Last-4 + Top-4 & 3.5\% & \textbf{19.1} & 32.5 & 27.3 & 29.3 & 13.5 & 53.7 & 77.0 & 36.1 \\
Last-8 + Top-8 & 7.0\% & 15.6 & 34.6 & 28.6 & 28.9 & 13.4 & 55.3 & 77.4 & 36.3 \\
\midrule
\multicolumn{10}{c}{\textbf{Segment-wise Compression}} \\
\midrule
Step-5  & 20.0\% & 16.8 & 34.2 & 31.2 & 32.1 & 14.2 & 65.5 & 79.7 & 39.1 \\
Step-10 & 10.2\% & 16.6 & 34.8 & 31.7 & 31.7 & \textbf{14.3} & 64.2 & \textbf{81.2} & \textbf{39.2} \\
Step-20 & 5.2\% & 16.1 & 34.5 & 32.0 & 31.3 & 14.0 & 61.9 & 79.2 & 38.4 \\
Step-30 & 3.5\% & 18.1 & 34.0 & 29.3 & 30.4 & 13.1 & 59.8 & 78.7 & 37.6 \\
\bottomrule
\end{tabular}
}
\caption{Results of different token compression strategies on seven BEIR datasets. We report NDCG@10 and the corresponding token retention ratio during inference.}
\label{tab:compression_ablation}
\end{table*}

\paragraph{Comparison with Reranking Baselines}

We further compare CompRank with representative retrieval and reranking baselines. 
Full results are provided in Appendix~\ref{app:complete_results}. 
BM25 and RankGPT obtain average NDCG@10 scores of 35.4 and 35.1. Attention-based rerankers achieve stronger results, with ICR (Mistral-7B), ICR (Qwen2.5-7B), and Corehead (Qwen2.5-7B) reaching 38.4, 38.3, and 38.1 average NDCG@10, respectively. CompRank achieves the best average score of 39.2 using the same Mistral-7B backbone as RankGPT and ICR, while also retaining only 10.2\% of document tokens. This shows that CompRank not only improves the efficiency of attention-based reranking, but also maintains competitive or better effectiveness compared with existing reranking methods.

%% file: sec/05_Analysis.tex
\section{Analysis}

\subsection{Efficiency Analysis}

We analyze the efficiency of CompRank from three aspects: block-level sparsity, token-level compression, and decoding-free scoring. First, CompRank removes document--document attention by encoding each document block independently and allowing the query block to attend to document blocks only during scoring. Compared with full attention over $N$ document blocks of length $L$, whose complexity scales as
$O((NL)^2d)$,
block decoupling reduces the document-side attention cost to:
$O(NL^2d)$,
changing the scaling with respect to the number of candidate documents from quadratic to linear.
Second, CompRank further reduces the query--document interaction cost by compressing the document-side KV states exposed to query-side attention. Let $\rho \in (0,1]$ denote the retained fraction of document tokens. Under a simplified setting where instruction, query, and document blocks have comparable length, an uncompressed block-wise reranker attends to instruction, query, and document tokens, with interaction cost proportional to $3L^2$. With token compression, only $\rho |D|$ document tokens are exposed to query-side attention, reducing the cost to approximately $(2+\rho)L^2$. The resulting attention-level speedup is:
\[
\mathcal{S}
\approx
\frac{3}{2+\rho}.
\]
With Segment-10 compression, $\rho \approx 0.1$, giving an estimated attention-level speedup of $1.43\times$. Empirically, Sec.~\ref{sec:scaling} shows that Segment-10 achieves approximately $1.3\times$ end-to-end speedup over the full-token CompRank variant.
Third, CompRank avoids autoregressive identifier generation by directly deriving document scores from attention distributions. This converts reranking from a sequential decoding process into a parallelizable scoring problem. As shown in Sec.~\ref{sec:scaling}, this decoding-free design achieves $4.9\times$--$9.5\times$ end-to-end speedup over a decode-ID listwise reranker when scaling from 30 to 500 candidate documents.
Finally, CompRank enables reusable document-side representations in principle. Since document blocks neither attend to other documents nor to the query block, their KV states can be pre-encoded or cached across repeated retrieval calls. This property is particularly useful in RAG and agentic search scenarios, where the same evidence documents may be revisited across multiple reasoning steps. Our current evaluation still recomputes document blocks, so the reported latency is a conservative end-to-end measurement without optimized KV caching. We provide a more detailed complexity and caching discussion in Appendix~\ref{app:efficiency_details}.

\subsection{Scaling and Extrapolation}
\label{sec:scaling}

\begin{figure*}[t]
    \centering
    \begin{subfigure}[b]{0.32\textwidth}
        \centering
        \includegraphics[width=\textwidth]{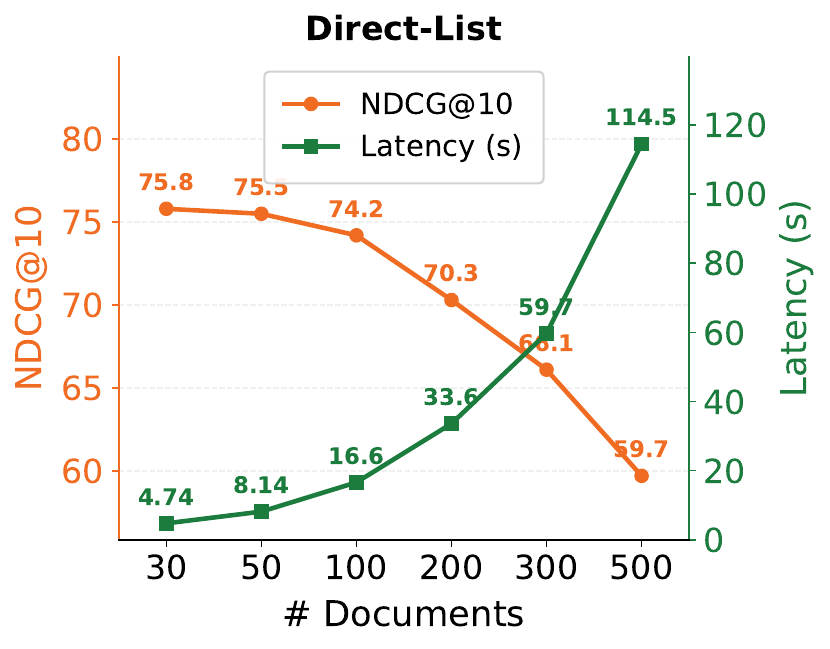}
    \end{subfigure}
    \hfill
    \begin{subfigure}[b]{0.32\textwidth}
        \centering
        \includegraphics[width=\textwidth]{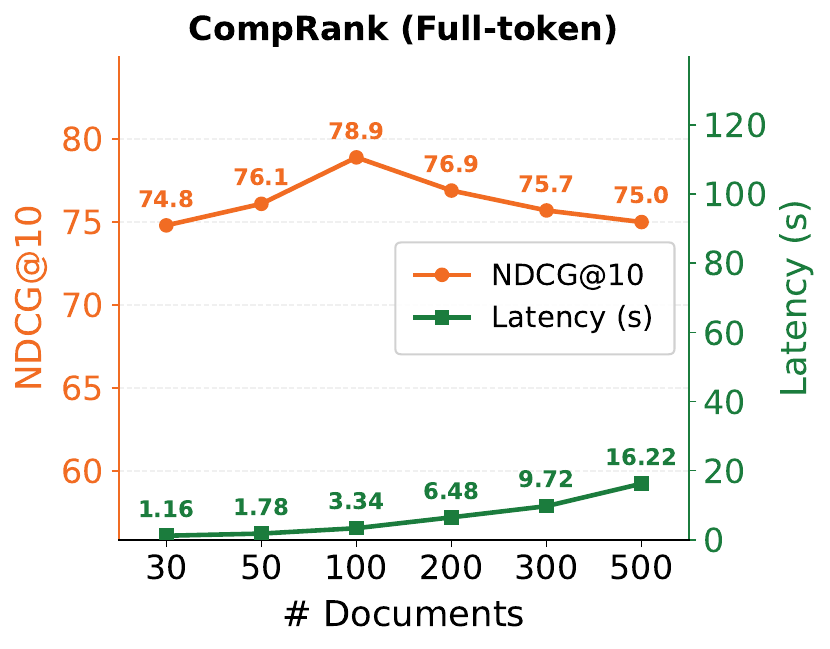}
    \end{subfigure}
    \hfill
    \begin{subfigure}[b]{0.32\textwidth}
        \centering
        \includegraphics[width=\textwidth]{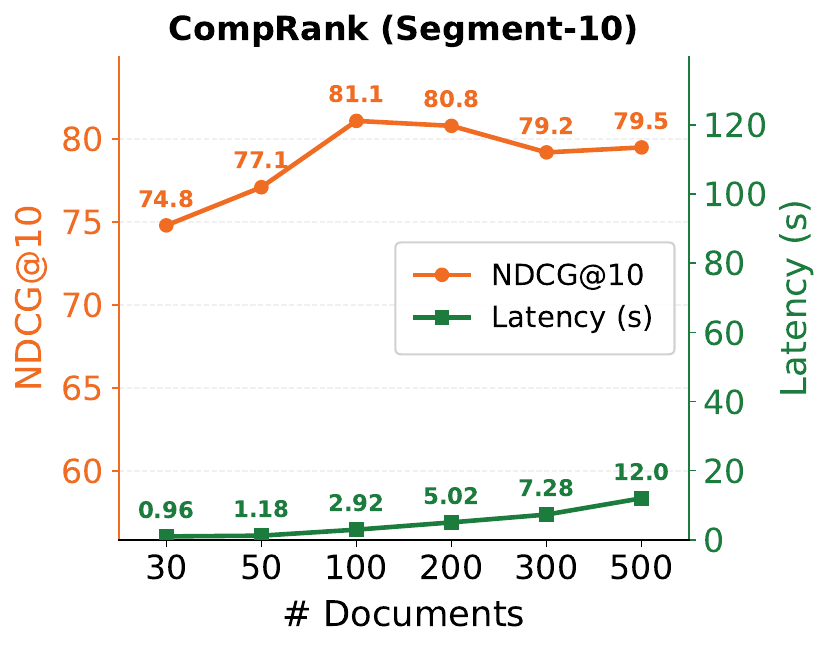}
    \end{subfigure}
    \caption{Scaling performance and end-to-end latency on TREC-COVID as the number of candidate documents increases. All models are trained on 30-document lists and evaluated up to 500 documents.}
    \label{fig:scaling}
\end{figure*}

We analyze the scalability and extrapolation behavior of CompRank under long candidate lists on the TREC-COVID dataset. Specifically, we compare CompRank with a generation-based Direct-List baseline that autoregressively decodes document identifiers. To separate the effect of token compression from attention-based scoring, we also include a full-token variant of CompRank. All models are trained on 30-document candidate lists and evaluated on larger candidate sets, ranging from 30 to 500 documents, without further fine-tuning. All experiments are conducted on a single A100 80GB GPU with batch size 1. Figure~\ref{fig:scaling} reports both NDCG@10 and end-to-end latency per query, showing a clear difference in extrapolation behavior.
Direct-List suffers a substantial performance drop as the candidate set grows, decreasing from 75.8 NDCG@10 at 30 documents to 59.7 at 500 documents. In contrast, both CompRank variants remain more stable. The full-token variant maintains competitive performance, while CompRank with Segment-10 reaches 79.5 NDCG@10 at 500 documents. This suggests that attention-based scoring is more robust than decode-ID generation in this long-list setting.

The degradation of Direct-List is likely related to its reliance on autoregressive decoding of document identifiers. Since the model is trained to generate identifiers within a limited range, increasing the number of candidates introduces distribution shift in the output space and may lead to invalid predictions, such as missing identifiers, duplicated identifiers, or out-of-range indices. A more detailed decoding error analysis is provided in Appendix~\ref{sec:appendix_decoding_errors}, where we observe that the model can maintain nearly error-free decoding when trained and evaluated on 30-document candidate sets, but the decoding error rate increases substantially when generalizing to larger candidate spaces (50--500 documents). This suggests that identifier-based autoregressive ranking struggles to generalize to unseen decoding sequences and larger output spaces.
CompRank also demonstrates substantially lower end-to-end latency. Direct-List becomes increasingly expensive as the candidate set grows, reaching 114.5s per query at 500 documents. In comparison, CompRank (Segment-10) requires only 12.0s, corresponding to a $9.5\times$ speedup. Across different candidate sizes, CompRank achieves $4.9\times$--$9.5\times$ speedup over Direct-List. Token-level compression further improves efficiency: compared with CompRank (Full-token), Segment-10 reduces latency from 16.22s to 12.0s at 500 documents, yielding approximately $1.3\times$ additional speedup while maintaining stronger ranking performance.

\paragraph{Latency breakdown.}
To further understand the source of efficiency gains, we decompose the latency of CompRank (Segment-10) into three components: end-to-end latency, full-forward latency, and query-block-forward latency. End-to-end latency measures the complete evaluation pipeline. Full-forward latency measures model forward computation when document blocks are recomputed for each query. Query-block-forward latency simulates a document pre-encoding setting, where document-side KV states are assumed to be cached and online computation only performs query-side forward computation and query-to-document scoring.

\begin{wraptable}{r}{0.48\textwidth}
\vspace{-1em}
\centering
\small
\setlength{\tabcolsep}{3pt}
\renewcommand{\arraystretch}{1.05}
\resizebox{0.46\textwidth}{!}{
\begin{tabular}{c|ccc}
\toprule
\textbf{\# Docs} 
& \textbf{End-to-End} 
& \textbf{Full-Forward} 
& \textbf{Query-Forward} \\
\midrule
30  & 0.96s  & 0.82s  & 0.016s \\
50  & 1.18s  & 1.02s  & 0.017s \\
100 & 2.92s  & 2.71s  & 0.022s \\
200 & 5.02s  & 4.76s  & 0.029s \\
300 & 7.28s  & 7.01s  & 0.037s \\
500 & 12.00s & 11.72s & 0.050s \\
\bottomrule
\end{tabular}
}
\caption{Latency breakdown of CompRank (Segment-10) on TREC-COVID. Query-forward simulates online computation with pre-encoded document-side KV states.}
\label{tab:latency_breakdown}
\vspace{-1em}
\end{wraptable}

As shown in Table~\ref{tab:latency_breakdown}, end-to-end latency is dominated by full model forward computation. When the number of candidate documents increases from 30 to 500, end-to-end latency grows from 0.96s/query to 12.00s/query, closely tracking full-forward latency, which grows from 0.82s/query to 11.72s/query. By contrast, the non-model overhead remains small, increasing only from roughly 0.14s/query to 0.28s/query. This indicates that the observed scaling cost mainly comes from recomputing document blocks rather than dataloading, padding, parsing, or metric computation.
Query-block-forward latency grows much more slowly, from 0.016s/query at 30 candidates to only 0.050s/query at 500 candidates. This highlights the potential benefit of document-side pre-encoding: if document KV states can be computed offline and reused during online reranking, the remaining online forward computation becomes substantially smaller. At 500 candidates, query-block-forward latency is $234\times$ lower than full-forward latency. We emphasize that this is a simulated pre-encoding analysis rather than a fully optimized serving implementation, since a real system must also account for KV-cache storage, loading, and memory-transfer overhead. Nevertheless, the breakdown suggests that CompRank's document-decoupled design provides a practical path toward low-latency online reranking when paired with system-level KV caching.

\subsection{Effect of CopyNet Supervision}

\begin{wrapfigure}{r}{0.48\textwidth}
\vspace{-1em}
\centering
\includegraphics[width=0.46\textwidth]{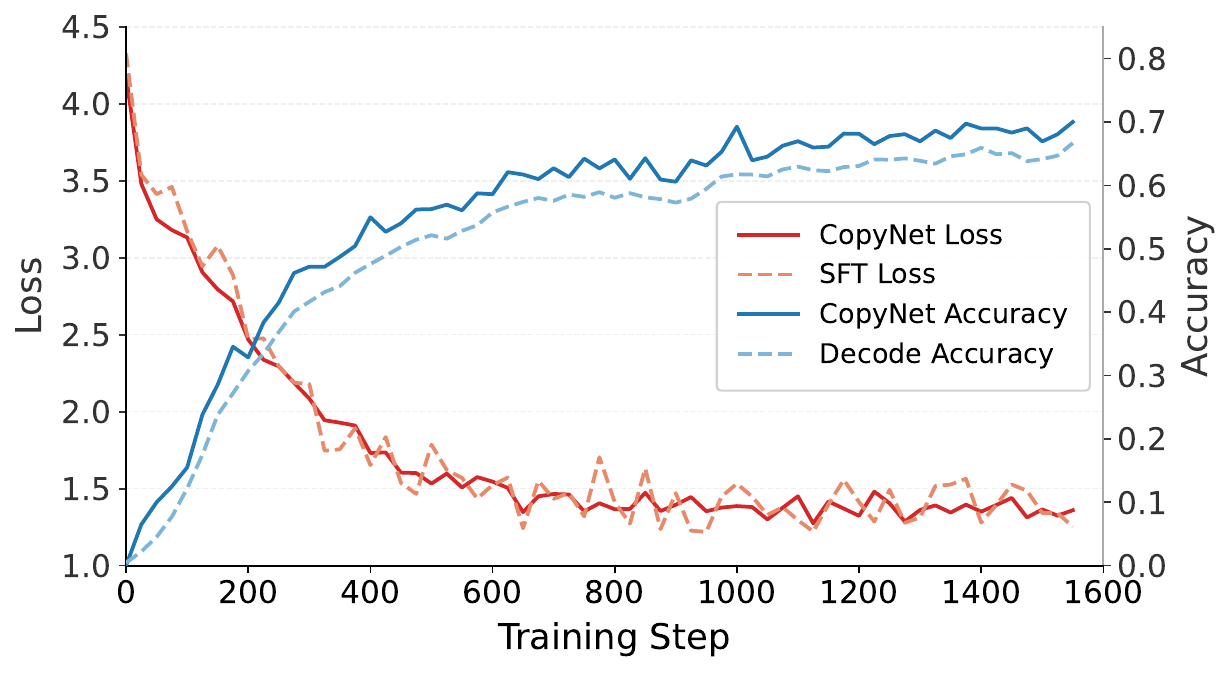}
\caption{Training dynamics of CopyNet and SFT supervision. CopyNet accuracy measures whether the positive document receives the highest attention-derived score, while decode accuracy measures whether SFT decoding predicts the positive identifier.}
\label{fig:copynet_training}
\vspace{-1em}
\end{wrapfigure}

We analyze whether CopyNet-style supervision provides an effective training signal for attention-based reranking, and compare it with a standard SFT objective that learns to generate the positive document identifier. The two objectives optimize different output spaces: SFT performs token-level language modeling over the full vocabulary, whereas CopyNet optimizes a pointer distribution over the input candidate documents using attention-derived scores. Therefore, the loss magnitudes are not directly comparable, but their training trends and the corresponding selection accuracies reveal how easily each objective aligns with document-level relevance.

Figure~\ref{fig:copynet_training} shows the training dynamics of both objectives. For CopyNet, we define accuracy as the proportion of training examples where the ground-truth positive document receives the highest attention-derived score among all candidates. For SFT, decode accuracy measures whether the decoded document identifier matches the ground-truth positive document. We observe that both CopyNet loss and SFT loss decrease steadily, suggesting that both objectives are learnable under the same training setting. However, CopyNet achieves consistently higher selection accuracy than SFT decoding, with CopyNet accuracy rising to around $70\%$ while decode accuracy remains slightly lower.
This behavior supports the use of CopyNet as an effective supervision signal for attention-based reranking. By restricting the prediction space to the input candidate documents, CopyNet directly supervises document-level scores rather than requiring the model to generate identifiers through the language modeling head. Since the same attention-derived scores are used for inference-time ranking, this objective naturally aligns training with the decoding-free scoring mechanism of CompRank. Overall, the results suggest that CopyNet supervision helps the model allocate attention-based scores toward relevant candidates and provides a direct training signal for document-level reranking.

%% file: sec/06_Conclusion.tex
\section{Conclusion}

We presented \textbf{CompRank}, a token-efficient reranking framework for long-list LLM-based reranking. CompRank reduces redundant computation by decoupling document representations, applying segment-wise token compression, and replacing autoregressive identifier decoding with attention-based scoring trained via a CopyNet-style objective.
Experiments on seven BEIR datasets show that segment-wise compression preserves most of the full-token reranking performance while retaining only 10.2\% of document tokens. The comparison across compression strategies further suggests that maintaining global document coverage is more effective than relying solely on final tokens or document-internal saliency estimates. Additional scaling experiments show that CompRank remains stable when evaluated on candidate lists up to 500 documents after training on 30-document lists, while substantially reducing end-to-end latency compared with generation-based listwise reranking.
Overall, our results suggest that token-level compression, document representation decoupling, and training--inference alignment provide a practical direction for scalable decoding-free LLM reranking. Future work includes implementing fully optimized document-side KV caching and exploring adaptive query-conditioned token selection.

%% file: sec/X_appendix.tex
\appendix
\label{sec:appendix}






\section{Implementation Details}
\label{app:implementation_details}

\subsection{CompRank Training Details}
\label{app:comprank_training_details}

We provide the implementation details used in our main CompRank experiments. The main results use Mistral-7B as the backbone model, with 30 candidate documents per query. CompRank derives document scores from attention logits at a fixed Transformer layer. In the Mistral-7B experiments, we use zero-based layer index 20, corresponding to the 21st Transformer block, as the scoring layer. The attention-based score is computed from the query positions in the final block. In our main setting, the effective decision-token set $U$ contains a single decision token, corresponding to the first supervised token in the completion, i.e., the token at the beginning of the answer sequence.
For each candidate document, CompRank first aggregates the softmax-normalized attention mass over the selected document tokens, and then averages the scores across attention heads. Since the main experiments use a single decision token, the aggregation operator $\operatorname{Agg}$ reduces to taking the score of this token. Our implementation also supports aggregation over multiple decision tokens using mean aggregation or log-sum-exp aggregation.

We train CompRank with a temperature-scaled CopyNet objective over candidate documents, using temperature $\tau=0.05$. The training objective is the CopyNet-style cross-entropy loss defined over attention-derived document scores, and no autoregressive SFT loss is used in the main experiments. Optimization uses AdamW 8-bit with a learning rate of $3.0\times10^{-6}$, weight decay 0, cosine learning-rate scheduling, and a warmup ratio of 0.03. Training is performed in bf16 precision with gradient checkpointing enabled for one epoch.
The training set contains 50,000 examples derived from the MS MARCO training split following the BlockRank data construction pipeline. We use a train--validation split ratio of 0.99, resulting in approximately 49,500 training examples and 500 validation examples. The per-device training batch size is 1 and the gradient accumulation step is 4, giving an effective batch size of $4 \times G$, where $G$ is the number of GPUs. Each candidate document block is truncated to at most 256 tokens, and each training instance contains 30 candidate documents.

\subsection{CompRank Input Prompt}
\label{app:comprank_prompt}

Fig.~\ref{fig:comprank_prompt} shows the prompt template for CompRank. Unlike Direct-List, CompRank does not require the model to autoregressively generate a complete ranked list of document identifiers. Instead, the prompt places candidate documents before the query and uses the beginning of the answer sequence as the decision position for attention-based scoring.

\begin{figure}[!htbp]
\centering
\begin{minipage}{0.95\linewidth}
\small
\begin{tcolorbox}[
    colback=gray!5,
    colframe=gray!50,
    title=\textbf{Input Prompt Template for CompRank},
    fonttitle=\bfseries,
    boxrule=0.5pt,
    arc=2pt,
    left=5pt,
    right=5pt,
    top=5pt,
    bottom=5pt
]
\textbf{Prompt:}
\begin{quote}
\ttfamily
You will be given a query and a list of documents. Each document will be formatted as CONTENT: <content> | END. Rank passages based on their relevance to the query.\\[0.3em]

Documents:\\
CONTENT: \{document\_0\} | END\\
CONTENT: \{document\_1\} | END\\
...\\
CONTENT: \{document\_N-1\} | END\\[0.3em]

====== Now let's start! ======\\
Find the most relevant document position.\\
Query: \{query\}\\
The following document(s) can help answer the query:\\[0.3em]

Final Answer:\\
\texttt{[}
\end{quote}
\end{tcolorbox}
\end{minipage}
\caption{Input prompt template used for CompRank. The decision token after \texttt{Final Answer:} is used for attention-based document scoring.}
\label{fig:comprank_prompt}
\end{figure}

\subsection{Direct-List Prompt and Decoding Setup}
\label{app:direct_list_prompt}

\begin{figure}[!htbp]
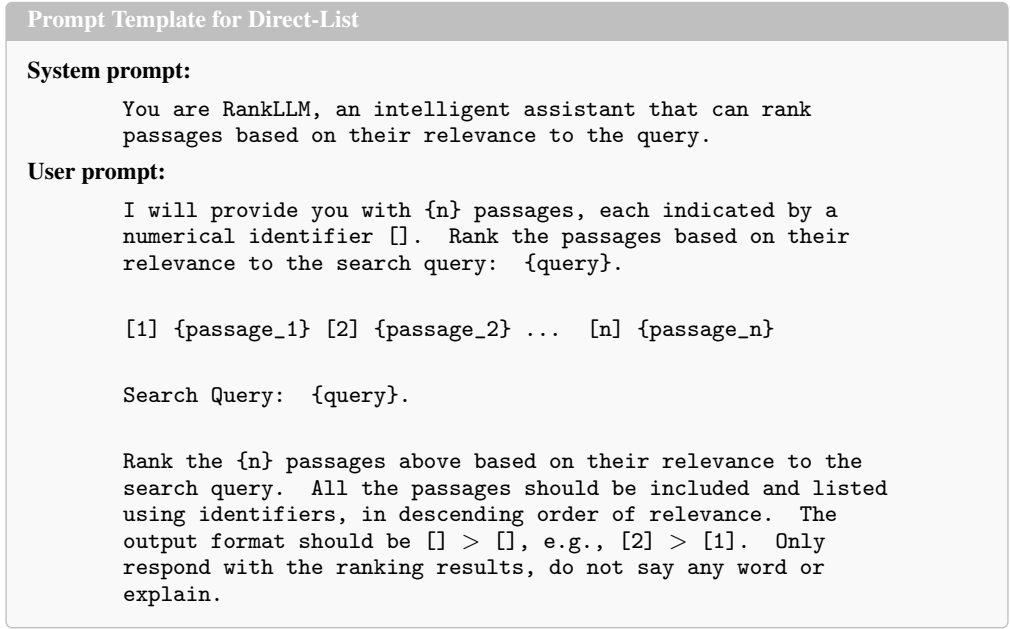

\centering
\begin{minipage}{0.95\linewidth}
\small
\begin{tcolorbox}[
    colback=gray!5,
    colframe=gray!50,
    title=\textbf{Prompt Template for Direct-List},
    fonttitle=\bfseries,
    boxrule=0.5pt,
    arc=2pt,
    left=5pt,
    right=5pt,
    top=5pt,
    bottom=5pt
]
\textbf{System prompt:}
\begin{quote}
\ttfamily
You are RankLLM, an intelligent assistant that can rank passages based on their relevance to the query.
\end{quote}

\textbf{User prompt:}
\begin{quote}
\ttfamily
I will provide you with \{n\} passages, each indicated by a numerical identifier []. Rank the passages based on their relevance to the search query: \{query\}.\\[0.3em]

[1] \{passage\_1\}
[2] \{passage\_2\}
...
[n] \{passage\_n\}\\[0.3em]

Search Query: \{query\}.\\[0.3em]

Rank the \{n\} passages above based on their relevance to the search query. All the passages should be included and listed using identifiers, in descending order of relevance. The output format should be [] $>$ [], e.g., [2] $>$ [1]. Only respond with the ranking results, do not say any word or explain.
\end{quote}
\end{tcolorbox}
\end{minipage}
\caption{Prompt template used for the Direct-List baseline.}
\label{fig:direct_list_prompt}
\end{figure}

Fig.~\ref{fig:direct_list_prompt} shows the prompt template for the Direct-List baseline. The model is instructed to generate a ranked list of document identifiers in descending order of relevance.

\section{Additional Efficiency Analysis}
\label{app:efficiency_details}

We provide a more detailed analysis of the efficiency properties of CompRank. The overall efficiency gain comes from four complementary design choices: block-level document decoupling, token-level compression, decoding-free scoring, and reusable document-side representations.

\paragraph{Structural sparsity via block decoupling.}
Standard listwise rerankers typically concatenate all candidate documents and apply full attention over the entire sequence. Let $N$ be the number of candidate documents, $L$ the document block length, and $d$ the hidden dimension. Full attention over the concatenated document sequence has complexity:
\[
O((NL)^2d).
\]
CompRank adopts a block-structured attention pattern that removes document--document interactions. Each document block is encoded independently, and the query block interacts with document blocks only during scoring. This reduces the document-side attention complexity to:
\[
O(NL^2d),
\]
which changes the scaling with respect to the number of documents from quadratic to linear. This structural sparsity is the basis for both more efficient computation and reusable document-side states.

\paragraph{Token-level compression.}
CompRank further reduces the query--document interaction cost by compressing document-side KV states. Let $\rho \in (0,1]$ denote the retained fraction of document tokens. For an uncompressed block-wise reranker, the query-side computation attends to instruction, query, and document tokens. Assuming for simplicity that $|I| \approx |Q| \approx |D| \approx L$, the interaction cost is proportional to:
\[
3L^2.
\]
With token compression, only $\rho |D|$ document tokens are exposed to query-side attention, reducing the cost to approximately:
\[
(2+\rho)L^2.
\]
The resulting attention-level speedup is:
\[
\mathcal{S}
=
\frac{|I|+|Q|+|D|}
{|I|+|Q|+\rho |D|}
\approx
\frac{3}{2+\rho}.
\]
With our default Segment-10 setting, $\rho \approx 0.1$, which gives:
\[
\frac{3}{2.1} \approx 1.43\times.
\]
This analysis estimates the reduction in attention-level computation. In practice, the observed wall-clock speedup can be smaller because the implementation may still include non-attention computation, kernel overhead, memory movement, and other evaluation overheads. In Sec.~\ref{sec:scaling}, Segment-10 yields approximately $1.3\times$ end-to-end speedup over the full-token CompRank variant, which is consistent with this theoretical estimate.

\paragraph{Decoding-free scoring.}
Generation-based listwise rerankers require autoregressive decoding to output ranked document identifiers. This introduces sequential generation overhead and may also lead to invalid outputs when the candidate list grows beyond the training setting. In contrast, CompRank derives document scores directly from attention distributions in a single forward pass and sorts the candidate documents according to these scores. This converts reranking from a generative decoding problem into a parallelizable scoring problem. As shown in Sec.~\ref{sec:scaling}, CompRank achieves $4.9\times$--$9.5\times$ end-to-end speedup over a decode-ID listwise reranker when scaling from 30 to 500 candidate documents.

\paragraph{Reusable document representations.}
CompRank also decouples document representations from both candidate order and query context. Since document blocks neither attend to other documents nor to the query block, their KV states can in principle be pre-encoded and reused:
\[
K_i, V_i = f_\theta(D_i).
\]
This enables a serving scenario where document-side computation is performed offline or cached across repeated retrieval calls, leaving only query-side computation and query-to-document attention on the online path. Such a design is particularly relevant for RAG and agentic search systems, where the same evidence documents may be retrieved or revisited across multiple reasoning steps.

We emphasize that our current evaluation pipeline still recomputes document blocks for each query. Therefore, the reported latency in Sec.~\ref{sec:scaling} should be interpreted as a conservative end-to-end measurement without optimized document-side KV caching. A production implementation with physical KV caching, token gathering, and efficient cache loading could further reduce online reranking cost by removing repeated document encoding from the critical path. At the same time, such a system would need to account for KV-cache storage, memory transfer, and cache lookup overhead, which we leave for future work.

\section{Complete Baseline Results}
\label{app:complete_results}

Table~\ref{tab:main_results_appendix} presents the complete per-dataset results for all retrieval and reranking baselines.

\begin{table*}[!htbp]
\centering
\small
\setlength{\tabcolsep}{5pt}
\renewcommand{\arraystretch}{1.15}
\resizebox{\textwidth}{!}{
\begin{tabular}{lcccccccc}
\toprule
\textbf{Model} 
& \textbf{FEVER} & \textbf{DBPedia} & \textbf{FiQA} & \textbf{NFCorpus} 
& \textbf{SciDocs} & \textbf{SciFact} & \textbf{COVID} & \textbf{Avg.} \\
\midrule
BM25 
& 16.5 & 31.8 & 23.6 & \textbf{33.8} & 14.9 & 67.9 & 59.5 & 35.4 \\
RankGPT (Mistral-7B) 
& 16.7 & 32.9 & 24.1 & 31.0 & 14.7 & 65.5 & 60.6 & 35.1 \\
ICR (Mistral-7B) 
& \textbf{20.6} & 31.4 & 31.0 & 33.2 & 16.2 & \textbf{72.4} & 63.9 & 38.4 \\
ICR (Qwen2.5-7B) 
& 19.6 & \textbf{35.3} & 27.0 & 32.7 & \textbf{16.4} & 71.1 & 66.1 & 38.3 \\
Corehead (Qwen2.5-7B) 
& 19.3 & 32.4 & 30.6 & 32.5 & 14.5 & 70.0 & 67.7 & 38.1 \\
\textbf{CompRank} (Mistral-7B) 
& 16.6 & 34.8 & \textbf{31.7} & 31.7 & 14.3 & 64.2 & \textbf{81.2} & \textbf{39.2} \\
\bottomrule
\end{tabular}
}
\caption{Complete reranking results on seven BEIR datasets. We report NDCG@10.}
\label{tab:main_results_appendix}
\end{table*}

\section{Ablation on Layer Selection and Decision Token Selection}
\subsection{Ablation on Transformer Layer Selection}
CompRank relies on the internal attention weights of the language model to score documents. Since different Transformer layers capture different levels of abstraction, the choice of layer significantly impacts the ranking performance. We evaluate the Attention Accuracy across all 32 layers of the Mistral-7B model, fixing the decision token to \texttt{'['}.

\begin{figure}[!htbp]
    \centering
    \includegraphics[width=0.8\textwidth]{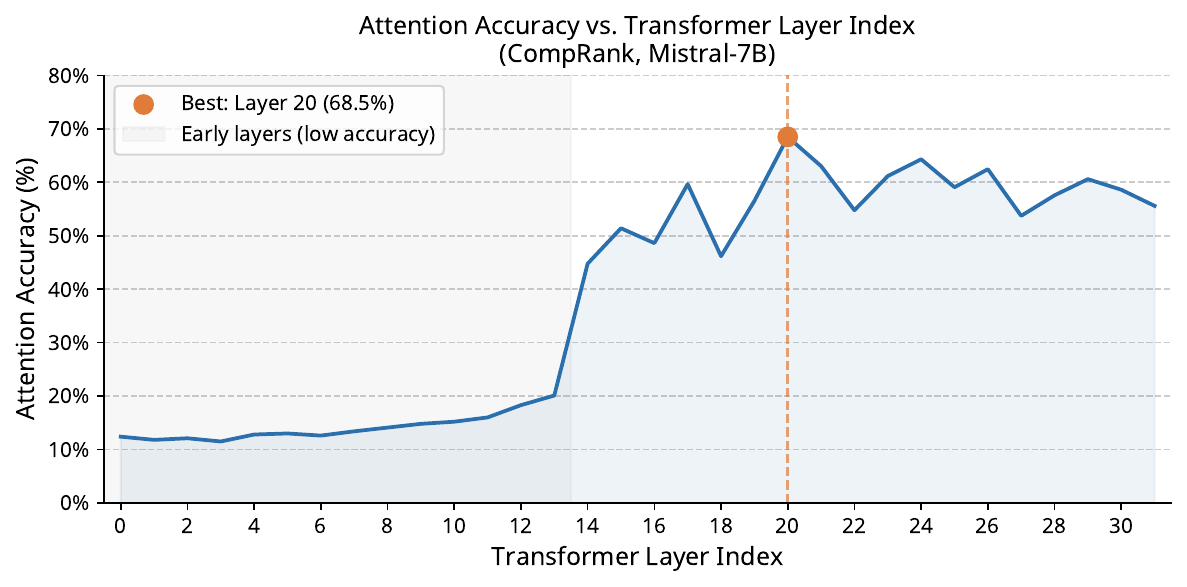}
    \caption{Attention Accuracy across all 32 Transformer layers of Mistral-7B. The metric measures the proportion of queries where the highest attention score is assigned to the positive document. The shaded region marks early layers with near-random performance.}
    \label{fig:layer_ablation}
\end{figure}

As illustrated in Figure~\ref{fig:layer_ablation}, the early layers (layers 0 to 13) exhibit very low Attention Accuracy (approximately 10--20\%), indicating that these layers have not yet aggregated sufficient semantic information to distinguish document relevance. Performance rises sharply around layer 14 and remains generally strong throughout the middle-to-late layers (layers 15 to 31), with natural oscillation. We observe the highest peak at zero-based layer index 20, corresponding to the 21st Transformer block, which achieves 68.5\% Attention Accuracy. Consequently, we select zero-based layer index 20, corresponding to the 21st Transformer block, as the default scoring layer for all Mistral-7B experiments.

\subsection{Ablation on Decision Token Selection}

The second critical design choice is determining which token(s) in the prompt or generation sequence should be used to aggregate the attention scores over candidate documents. We compare the following five configurations, all evaluated at layer 20:

\begin{itemize}
    \item \textbf{Query Token Aggregation:} Attention scores are aggregated over the individual query token positions only (excluding non-query tokens).
    \item \textbf{Query Block Aggregation:} Attention scores are averaged over all tokens in the entire query block.
    \item \textbf{\texttt{'Final'}, \texttt{'Answer'}:} Individual tokens from the answer-prefix template are used as the decision token.
    \item \textbf{\texttt{'['}:} The first generated token of the output sequence, which opens the listwise ranking output.
\end{itemize}

\begin{figure}[h]
    \centering
    \includegraphics[width=0.72\textwidth]{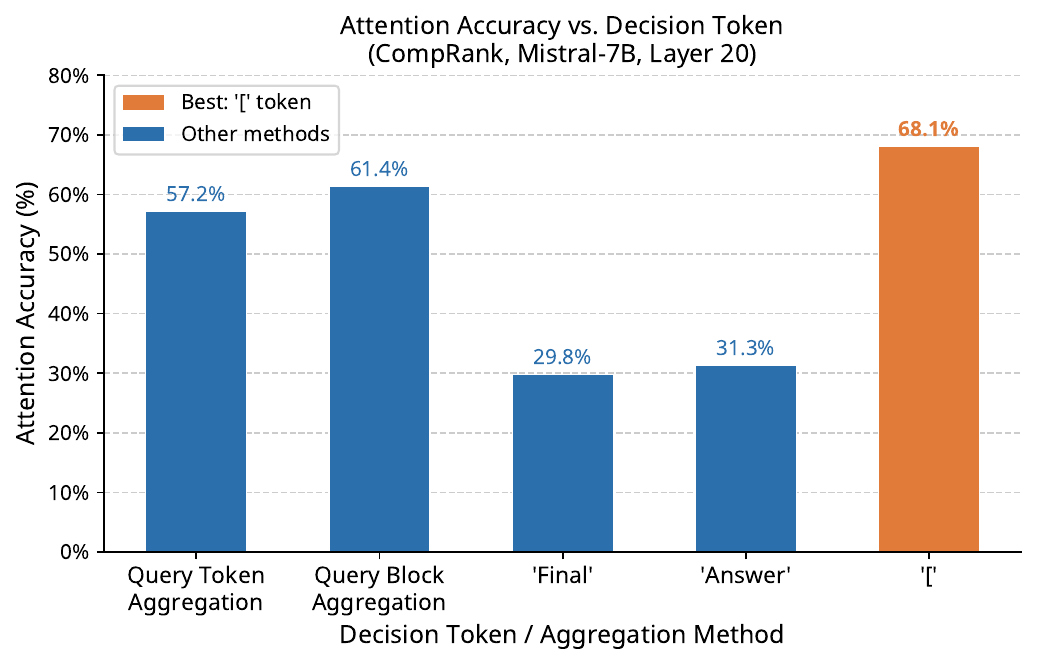}
    \caption{Attention Accuracy under different decision token and aggregation strategies (CompRank, Mistral-7B, Layer 20). Query Token Aggregation and Query Block Aggregation aggregate attention over multiple positions, while the remaining three use a single token as the decision point.}
    \label{fig:token_ablation}
\end{figure}

Figure~\ref{fig:token_ablation} shows the Attention Accuracy for each configuration. Query Token Aggregation and Query Block Aggregation achieve 57.2\% and 61.4\% respectively, demonstrating that aggregating over query-side positions provides a reasonable signal. However, the answer-prefix tokens \texttt{'Final'} and \texttt{'Answer'} perform substantially worse (29.8\% and 31.3\%), suggesting that these positions do not concentrate relevance-discriminative attention. The highest Attention Accuracy of 68.1\% is achieved by using the first generated token \texttt{'['} as the sole decision token. This token marks the exact moment the model begins producing its ranking output, and its attention distribution appears to be the most focused and informative for document relevance scoring. We therefore adopt \texttt{'['} as the default decision token in all CompRank experiments.

\section{Analysis of Decoding Errors in Direct-List}
\label{sec:appendix_decoding_errors}

In Section~\ref{sec:scaling}, we observe that the generation-based Direct-List baseline degrades as the candidate list becomes longer. To better understand this behavior, we analyze its decoded identifier sequences on TREC-COVID when the number of candidate documents increases from 30 to 500. Since the Direct-List model is trained with 30-document candidate lists, we first examine whether it can correctly produce valid identifier sequences under the same list size, and then evaluate how the decoding behavior changes under longer candidate lists.
We consider four types of decoding errors:
\begin{itemize}
    \item \textbf{Duplicate IDs:} occur when the same document identifier is generated multiple times within a single ranked list. 
    \item \textbf{Out-of-range IDs:} occur when the model generates an identifier outside the valid candidate range. 
    \item \textbf{Missing IDs:} refer to valid candidate identifiers that are not generated in the output sequence. 
    \item \textbf{Truncated Output:} occurs when the model terminates generation before producing the expected number of identifiers. 
\end{itemize}

These errors directly affect listwise ranking quality because the generated sequence no longer corresponds to a valid permutation of the candidate set.

\begin{table}[h]
\centering
\caption{Decoding error statistics of Direct-List on TREC-COVID (50 queries). Percentages indicate the proportion of queries exhibiting each error type.}
\label{tab:error_stats}
\resizebox{\textwidth}{!}{
\begin{tabular}{cccccc}
\toprule
\textbf{List Size} & \textbf{Any Error} & \textbf{Duplicate IDs} & \textbf{Out-of-Range IDs} & \textbf{Missing IDs} & \textbf{Truncated Output} \\
\midrule
30  & 0 (0.0\%)    & 0 (0.0\%)   & 0 (0.0\%) & 0 (0.0\%)   & 0 (0.0\%) \\
50  & 46 (92.0\%)  & 17 (34.0\%) & 0 (0.0\%) & 43 (86.0\%) & 32 (64.0\%) \\
100 & 47 (94.0\%)  & 21 (42.0\%) & 3 (6.0\%) & 46 (92.0\%) & 33 (66.0\%) \\
200 & 50 (100.0\%) & 29 (58.0\%) & 0 (0.0\%) & 49 (98.0\%) & 36 (72.0\%) \\
300 & 50 (100.0\%) & 31 (62.0\%) & 2 (4.0\%) & 48 (96.0\%) & 43 (86.0\%) \\
400 & 50 (100.0\%) & 39 (78.0\%) & 0 (0.0\%) & 50 (100.0\%)& 36 (72.0\%) \\
500 & 50 (100.0\%) & 33 (66.0\%) & 0 (0.0\%) & 50 (100.0\%)& 41 (82.0\%) \\
\bottomrule
\end{tabular}
}
\end{table}

\begin{figure}[t]
    \centering
    \includegraphics[width=0.78\textwidth]{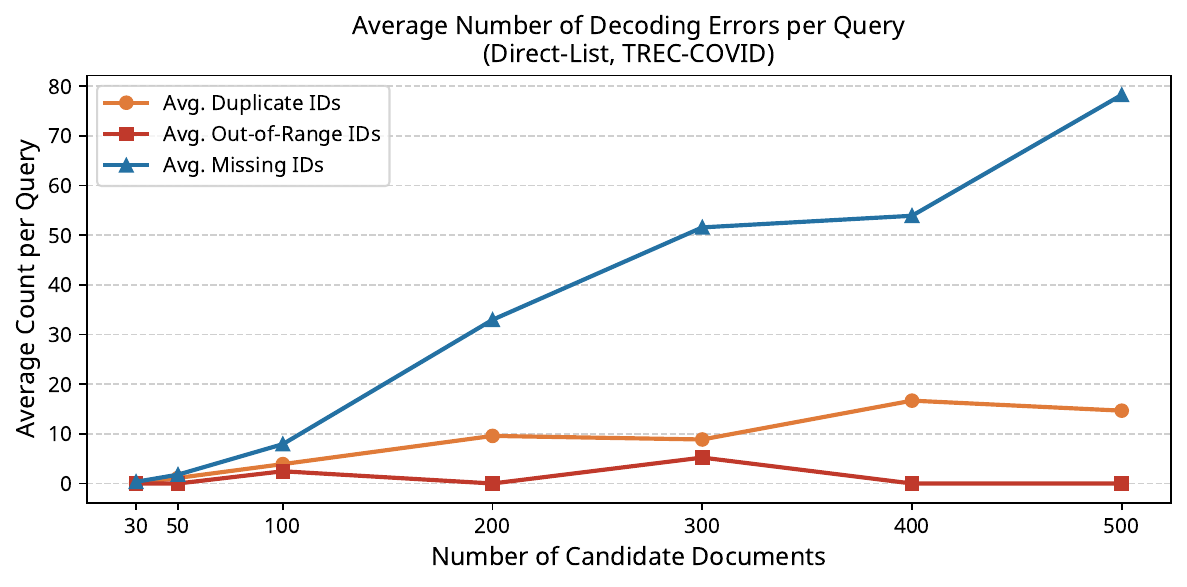}
    \caption{Average number of decoding errors per query for Direct-List on TREC-COVID as the number of candidate documents increases. The model produces valid identifier sequences under the 30-document setting seen during training, but decoding errors increase when evaluated on longer candidate lists.}
    \label{fig:avg_decoding_errors}
\end{figure}

Table~\ref{tab:error_stats} reports the proportion of queries affected by each error type, while Figure~\ref{fig:avg_decoding_errors} shows the average number of errors per query. When evaluated with 30 candidates, which matches the training list size, Direct-List produces valid outputs without duplicate, out-of-range, missing, or truncated identifiers. However, once the candidate list grows beyond the training setting, decoding errors appear rapidly. At 50 candidates, 92.0\% of queries already contain at least one error; from 200 candidates onward, all queries exhibit decoding errors.
Missing identifiers are the dominant failure mode. As shown in Table~\ref{tab:error_stats}, the proportion of queries with missing IDs increases from 86.0\% at 50 candidates to 100.0\% at 400 and 500 candidates. Figure~\ref{fig:avg_decoding_errors} further shows that the average number of missing IDs per query grows substantially, reaching 78.2 under the 500-document setting. Duplicate IDs also become frequent, affecting 66.0\% of queries at 500 candidates, indicating that the model struggles to maintain a valid one-to-one mapping between generation steps and candidate identifiers. Out-of-range IDs occur less frequently, but their presence shows that the generated sequence is not always constrained to the valid candidate space.
Overall, these results suggest that decode-ID listwise reranking has limited extrapolation ability when the inference-time candidate list is substantially longer than the training-time list. Although the model can generate valid identifier sequences for the 30-document setting it was trained on, extending this pattern to 50--500 documents introduces a distribution shift in the output sequence space. As a result, the model increasingly fails to produce a valid permutation of candidate identifiers, providing additional motivation for CompRank's decoding-free attention-based scoring paradigm.

\begin{figure}[t]
\centering
\begin{minipage}{0.95\linewidth}
\small
\begin{tcolorbox}[
    colback=gray!5,
    colframe=gray!50,
    title=\textbf{Case Study: Duplicate and Missing IDs in Decode-ID Reranking},
    fonttitle=\bfseries,
    boxrule=0.5pt,
    arc=2pt,
    left=4pt,
    right=4pt,
    top=4pt,
    bottom=4pt
]
\textbf{Dataset:} TREC-COVID \\
\textbf{Query:} \textit{How does the coronavirus differ from seasonal flu?} \\
\textbf{Candidate List Size:} 50 \\
\textbf{Expected Output Length:} 50 \\
\textbf{Actual Output Length:} 56 \\[0.3em]

\textbf{Duplicate IDs:} \texttt{[3, 4, 12, 13, 17, 19, 26, 27, 35]} \\
\textbf{Missing IDs:} \texttt{[29, 33, 39]} \\[0.3em]

\textbf{Raw Output Snippet:}
\begin{quote}
\ttfamily
[1] $>$ [21] $>$ [22] $>$ [20] $>$ [5] $>$ [13] $>$ [26] $>$ [27] $>$ [35] $>$ [4] $>$ [17] $>$ [19] $>$ [3] $>$ [12] $>$ [24] $>$ [18] $>$ [6] $>$ [16] $>$ [25] $>$ [42] $>$ [43] $>$ [7] $>$ [9] $>$ [14] $>$ [8] $>$ [31] $>$ [32] $>$ [28] $>$ [34] $>$ [44] $>$ [48] $>$ [10] $>$ [11] $>$ [15] $>$ [23] $>$ [30] $>$ [36] $>$ [37] $>$ [38] $>$ [40] $>$ [41] $>$ [45] $>$ [46] $>$ [47] $>$ [49] $>$ [50] $>$ [2] $>$ [13] $>$ [26] $>$ [27] $>$ [35] $>$ [4] $>$ [17] $>$ [19] $>$ [3] $>$ [12] \texttt{<|im\_end|>}
\end{quote}

\textbf{Observation:} The model initially generates many valid identifiers, but later repeats a block of previously generated IDs. This creates both duplicate IDs and missing candidate IDs, so the output no longer forms a valid permutation of the 50 candidate documents.
\end{tcolorbox}
\end{minipage}
\caption{Qualitative example of decode-ID failure in Direct-List. The model repeats previously generated identifiers and omits several valid candidates when evaluated beyond its 30-document training setting.}
\label{fig:decode_id_case_study_50}
\end{figure}

\begin{figure}[t]
\centering
\begin{minipage}{0.95\linewidth}
\small
\begin{tcolorbox}[
    colback=gray!5,
    colframe=gray!50,
    title=\textbf{Case Study: Severe Coverage Failure under Long Candidate Lists},
    fonttitle=\bfseries,
    boxrule=0.5pt,
    arc=2pt,
    left=4pt,
    right=4pt,
    top=4pt,
    bottom=4pt
]
\textbf{Dataset:} TREC-COVID \\
\textbf{Query:} \textit{What are the mortality rates overall and in specific populations?} \\
\textbf{Candidate List Size:} 300 \\
\textbf{Expected Output Length:} 300 \\
\textbf{Actual Output Length:} 295 \\[0.3em]

\textbf{Duplicate IDs:} \texttt{[192, 193, 194, ..., 201]} \\
\textbf{Missing IDs:} \texttt{[16, 17, 18, 19, 21, 22, 23, 24, 26, 27, ...]} \\[0.3em]

\textbf{Observation:} The model omits a large number of valid identifiers and repeatedly generates a subset of IDs. This indicates that identifier generation becomes increasingly unstable when the output sequence is much longer than those seen during training.
\end{tcolorbox}
\end{minipage}
\caption{Qualitative example of severe coverage failure for Direct-List under a 300-document candidate list.}
\label{fig:decode_id_case_study_300}
\end{figure}

\section{Limitations}
\label{app:limitations}

CompRank has several limitations. First, our main experiments use Mistral-7B as the primary backbone, and broader validation across more model families remains future work. Second, although document representation decoupling enables reusable document-side KV states in principle, our current evaluation pipeline still recomputes document blocks and does not implement a fully optimized KV-caching system.  Third, static segment-wise compression may be suboptimal when relevance evidence is sparse or highly localized, where adaptive query-conditioned token selection could be more effective. Finally, our results are based on single-run experiments due to computational constraints, and future work should include more extensive statistical evaluation.

%% file: neurips_2026.bib
@article{rag,
  title={Retrieval-augmented generation for knowledge-intensive nlp tasks},
  author={Lewis, Patrick and Perez, Ethan and Piktus, Aleksandra and Petroni, Fabio and Karpukhin, Vladimir and Goyal, Naman and K{\"u}ttler, Heinrich and Lewis, Mike and Yih, Wen-tau and Rockt{\"a}schel, Tim and others},
  journal={Advances in neural information processing systems},
  volume={33},
  pages={9459--9474},
  year={2020}
}

@inproceedings{
lu2026rethinking,
title={Rethinking Reasoning in Document Ranking: Why Chain-of-Thought Falls Short},
author={Xuan Lu and Haohang Huang and Rui Meng and Yaohui Jin and Wenjun Zeng and Xiaoyu Shen},
booktitle={The Fourteenth International Conference on Learning Representations},
year={2026},
url={https://openreview.net/forum?id=txmqENuRcc}
}

@inproceedings{lumulticonir,
    title = "{M}ulti{C}on{IR}: Towards Multi-Condition Information Retrieval",
    author = "Lu, Xuan  and
      Liu, Sifan  and
      Yin, Bochao  and
      Li, Yongqi  and
      Chen, Xinghao  and
      Su, Hui  and
      Jin, Yaohui  and
      Zeng, Wenjun  and
      Shen, Xiaoyu",
    editor = "Christodoulopoulos, Christos  and
      Chakraborty, Tanmoy  and
      Rose, Carolyn  and
      Peng, Violet",
    booktitle = "Findings of the Association for Computational Linguistics: EMNLP 2025",
    month = nov,
    year = "2025",
    address = "Suzhou, China",
    publisher = "Association for Computational Linguistics",
    url = "https://aclanthology.org/2025.findings-emnlp.726/",
    pages = "13471--13494",
    ISBN = "979-8-89176-335-7"
}

@inproceedings{
lu2026tools,
title={Tools are under-documented: Simple Document Expansion Boosts Tool Retrieval},
author={Xuan Lu and Haohang Huang and Rui Meng and Yaohui Jin and Wenjun Zeng and Xiaoyu Shen},
booktitle={The Fourteenth International Conference on Learning Representations},
year={2026},
url={https://openreview.net/forum?id=g9D9MgG7iW}
}

@inproceedings{lu2026beyond,
  title={Beyond Global Similarity: Multi-Conditional Retrieval for Fine-Grained Cross-Modal Understanding},
  author={Lu, Xuan and Li, Kangle and Huang, Haohang and Meng, Rui and Zeng, Wenjun and Shen, Xiaoyu},
  booktitle={Proceedings of the IEEE/CVF Conference on Computer Vision and Pattern Recognition},
  pages={9699--9709},
  year={2026}
}

@article{huang2026mmeb,
  title={MMEB-V3: Measuring the Performance Gaps of Omni-Modality Embedding Models},
  author={Huang, Haohang and Lu, Xuan and Su, Mingyi and Zhang, Xuan and Jiang, Ziyan and Nie, Ping and Zou, Kai and Pfister, Tomas and Chen, Wenhu and Zhang, Wei and others},
  journal={arXiv preprint arXiv:2604.23321},
  year={2026}
}

@misc{yin2026queriesdecomposedstageawarestudy,
      title={When Should Queries Be Decomposed? A Stage-Aware Study of Query Decomposition for Multi-Condition Retrieval}, 
      author={Bochao Yin and Xuan Lu and Zhengyu Qi and Xiaoyu Shen},
      year={2026},
      eprint={2606.08577},
      archivePrefix={arXiv},
      primaryClass={cs.IR},
      url={https://arxiv.org/abs/2606.08577}, 
}

@misc{fan2026minirerankerefficientmultimodalreranking,
      title={miniReranker: Efficient Multimodal Reranking through Visual Cache Reuse and Interaction Sparsity}, 
      author={Yingqi Fan and Xuan Lu and Anhao Zhao and Junlong Tong and Ping Nie and Kai Zou and Yunpu Ma and Wei Zhang and Xiaoyu Shen},
      year={2026},
      eprint={2606.10759},
      archivePrefix={arXiv},
      primaryClass={cs.IR},
      url={https://arxiv.org/abs/2606.10759}, 
}

@inproceedings{qin-etal-2024-large,
    title = "Large Language Models are Effective Text Rankers with Pairwise Ranking Prompting",
    author = "Qin, Zhen  and
      Jagerman, Rolf  and
      Hui, Kai  and
      Zhuang, Honglei  and
      Wu, Junru  and
      Yan, Le  and
      Shen, Jiaming  and
      Liu, Tianqi  and
      Liu, Jialu  and
      Metzler, Donald  and
      Wang, Xuanhui  and
      Bendersky, Michael",
    editor = "Duh, Kevin  and
      Gomez, Helena  and
      Bethard, Steven",
    booktitle = "Findings of the Association for Computational Linguistics: NAACL 2024",
    month = jun,
    year = "2024",
    address = "Mexico City, Mexico",
    publisher = "Association for Computational Linguistics",
    url = "https://aclanthology.org/2024.findings-naacl.97/",
    doi = "10.18653/v1/2024.findings-naacl.97",
    pages = "1504--1518"
}

@misc{pradeep2023rankvicunazeroshotlistwisedocument,
      title={RankVicuna: Zero-Shot Listwise Document Reranking with Open-Source Large Language Models}, 
      author={Ronak Pradeep and Sahel Sharifymoghaddam and Jimmy Lin},
      year={2023},
      eprint={2309.15088},
      archivePrefix={arXiv},
      primaryClass={cs.IR},
      url={https://arxiv.org/abs/2309.15088}, 
}

@inproceedings{sun2023rankgpt,
  title={Is ChatGPT Good at Search? Investigating Large Language Models as Re-Ranking Agents},
  author={Sun, Weiwei and Yan, Lingyong and Ma, Xinyu and Wang, Shuaiqiang and Ren, Pengjie and Chen, Zhumin and Yin, Dawei and Ren, Zhaochun},
  booktitle={Proceedings of the 2023 Conference on Empirical Methods in Natural Language Processing},
  pages={14918--14937},
  year={2023}
}

@misc{pradeep2023rankzephyreffectiverobustzeroshot,
      title={RankZephyr: Effective and Robust Zero-Shot Listwise Reranking is a Breeze!}, 
      author={Ronak Pradeep and Sahel Sharifymoghaddam and Jimmy Lin},
      year={2023},
      eprint={2312.02724},
      archivePrefix={arXiv},
      primaryClass={cs.IR},
      url={https://arxiv.org/abs/2312.02724}, 
}

@inproceedings{zhuang2024setwise,
  title={A setwise approach for effective and highly efficient zero-shot ranking with large language models},
  author={Zhuang, Shengyao and Zhuang, Honglei and Koopman, Bevan and Zuccon, Guido},
  booktitle={Proceedings of the 47th International ACM SIGIR Conference on Research and Development in Information Retrieval},
  pages={38--47},
  year={2024}
}

@inproceedings{sharifymoghaddam2025rankllm,
  title={RankLLM: A Python Package for Reranking with LLMs},
  author={Sharifymoghaddam, Sahel and Pradeep, Ronak and Slavescu, Andre and Nguyen, Ryan and Xu, Andrew and Chen, Zijian and Zhang, Yilin and Chen, Yidi and Xian, Jasper and Lin, Jimmy},
  booktitle={Proceedings of the 48th International ACM SIGIR Conference on Research and Development in Information Retrieval},
  pages={3681--3690},
  year={2025}
}

@inproceedings{monot5,
  title={Document Ranking with a Pretrained Sequence-to-Sequence Model},
  author={Nogueira, Rodrigo Frassetto and Jiang, Zhiying and Pradeep, Ronak and Lin, Jimmy},
  booktitle={EMNLP (Findings)},
  year={2020}
}

@inproceedings{rankllama,
  title={Fine-tuning llama for multi-stage text retrieval},
  author={Ma, Xueguang and Wang, Liang and Yang, Nan and Wei, Furu and Lin, Jimmy},
  booktitle={Proceedings of the 47th International ACM SIGIR Conference on Research and Development in Information Retrieval},
  pages={2421--2425},
  year={2024}
}

@misc{monobert,
      title={Multi-Stage Document Ranking with BERT}, 
      author={Rodrigo Nogueira and Wei Yang and Kyunghyun Cho and Jimmy Lin},
      year={2019},
      eprint={1910.14424},
      archivePrefix={arXiv},
      primaryClass={cs.IR},
      url={https://arxiv.org/abs/1910.14424}, 
}

@inproceedings{
blockrank,
title={Scalable In-context Ranking with Generative Models},
author={Nilesh Gupta and Chong You and Srinadh Bhojanapalli and Sanjiv Kumar and Inderjit S Dhillon and Felix X. Yu},
booktitle={The Thirty-ninth Annual Conference on Neural Information Processing Systems},
year={2025},
url={https://openreview.net/forum?id=zj45hoQhjD}
}

@inproceedings{
chen2025attention,
title={Attention in Large Language Models Yields Efficient Zero-Shot Re-Rankers},
author={Shijie Chen and Bernal Jimenez Gutierrez and Yu Su},
booktitle={The Thirteenth International Conference on Learning Representations},
year={2025},
url={https://openreview.net/forum?id=yzloNYH3QN}
}

@inproceedings{QRHead,
    title = "Query-Focused Retrieval Heads Improve Long-Context Reasoning and Re-ranking",
    author = "Zhang, Wuwei  and
      Yin, Fangcong  and
      Yen, Howard  and
      Chen, Danqi  and
      Ye, Xi",
    editor = "Christodoulopoulos, Christos  and
      Chakraborty, Tanmoy  and
      Rose, Carolyn  and
      Peng, Violet",
    booktitle = "Proceedings of the 2025 Conference on Empirical Methods in Natural Language Processing",
    month = nov,
    year = "2025",
    address = "Suzhou, China",
    publisher = "Association for Computational Linguistics",
    url = "https://aclanthology.org/2025.emnlp-main.1214/",
    doi = "10.18653/v1/2025.emnlp-main.1214",
    pages = "23791--23805",
    ISBN = "979-8-89176-332-6"
}

@inproceedings{zhou2026longranker,
  title={LongRanker: Efficient One-Pass Document Reranking with Long-Context Large Language Models},
  author={Zhou, Changjiang and Zhang, Ruqing and Guo, Jiafeng and de Rijke, Maarten and Yixing, Fan and Cheng, Xueqi},
  booktitle={Proceedings of the ACM Web Conference 2026},
  pages={2004--2013},
  year={2026}
}

@misc{hyperRAG,
      title={HyperRAG: Enhancing Quality-Efficiency Tradeoffs in Retrieval-Augmented Generation with Reranker KV-Cache Reuse}, 
      author={Yuwei An and Yihua Cheng and Seo Jin Park and Junchen Jiang},
      year={2025},
      eprint={2504.02921},
      archivePrefix={arXiv},
      primaryClass={cs.CL},
      url={https://arxiv.org/abs/2504.02921}, 
}

@misc{dejean2025rerankingcompresseddocumentrepresentation,
      title={Reranking with Compressed Document Representation}, 
      author={Hervé Déjean and Stéphane Clinchant},
      year={2025},
      eprint={2505.15394},
      archivePrefix={arXiv},
      primaryClass={cs.IR},
      url={https://arxiv.org/abs/2505.15394}, 
}

@misc{li2026efficientlongdocumentrerankingblocklevel,
      title={Efficient Long-Document Reranking via Block-Level Embeddings and Top-k Interaction Refinement}, 
      author={Minghan Li and Eric Gaussier and Guodong Zhou},
      year={2026},
      eprint={2501.17039},
      archivePrefix={arXiv},
      primaryClass={cs.IR},
      url={https://arxiv.org/abs/2501.17039}, 
}

@misc{wang2026headrankdecodingfreepassagereranking,
      title={HeadRank: Decoding-Free Passage Reranking via Preference-Aligned Attention Heads}, 
      author={Juyuan Wang and Chenxing Wang and Yuchen Fang and Huiyun Hu and Junwu Du and Aolin Li and Haijun Wu and Jin Xu and Ligang Liu and Dongliang Liao},
      year={2026},
      eprint={2604.17237},
      archivePrefix={arXiv},
      primaryClass={cs.IR},
      url={https://arxiv.org/abs/2604.17237}, 
}

@misc{tran2026contrastiveretrievalheadsimprove,
      title={Contrastive Retrieval Heads Improve Attention-Based Re-Ranking}, 
      author={Linh Tran and Yulong Li and Radu Florian and Wei Sun},
      year={2026},
      eprint={2510.02219},
      archivePrefix={arXiv},
      primaryClass={cs.IR},
      url={https://arxiv.org/abs/2510.02219}, 
}

@inproceedings{copynet,
    title = "Incorporating Copying Mechanism in Sequence-to-Sequence Learning",
    author = "Gu, Jiatao  and
      Lu, Zhengdong  and
      Li, Hang  and
      Li, Victor O.K.",
    editor = "Erk, Katrin  and
      Smith, Noah A.",
    booktitle = "Proceedings of the 54th Annual Meeting of the Association for Computational Linguistics (Volume 1: Long Papers)",
    month = aug,
    year = "2016",
    address = "Berlin, Germany",
    publisher = "Association for Computational Linguistics",
    url = "https://aclanthology.org/P16-1154/",
    doi = "10.18653/v1/P16-1154",
    pages = "1631--1640"
}

@inproceedings{
    thakur2021beir,
    title={{BEIR}: A Heterogeneous Benchmark for Zero-shot Evaluation of Information Retrieval Models},
    author={Nandan Thakur and Nils Reimers and Andreas R{\"u}ckl{\'e} and Abhishek Srivastava and Iryna Gurevych},
    booktitle={Thirty-fifth Conference on Neural Information Processing Systems Datasets and Benchmarks Track (Round 2)},
    year={2021},
    url={https://openreview.net/forum?id=wCu6T5xFjeJ}
}

@misc{beltagy2020longformer,
      title={Longformer: The Long-Document Transformer}, 
      author={Iz Beltagy and Matthew E. Peters and Arman Cohan},
      year={2020},
      eprint={2004.05150},
      archivePrefix={arXiv},
      primaryClass={cs.CL},
      url={https://arxiv.org/abs/2004.05150}, 
}

@article{zaheer2020bigbird,
  title={Big bird: Transformers for longer sequences},
  author={Zaheer, Manzil and Guruganesh, Guru and Dubey, Kumar Avinava and Ainslie, Joshua and Alberti, Chris and Ontanon, Santiago and Pham, Philip and Ravula, Anirudh and Wang, Qifan and Yang, Li and others},
  journal={Advances in neural information processing systems},
  volume={33},
  pages={17283--17297},
  year={2020}
}

@misc{rankgpt,
      title={Is ChatGPT Good at Search? Investigating Large Language Models as Re-Ranking Agents}, 
      author={Weiwei Sun and Lingyong Yan and Xinyu Ma and Shuaiqiang Wang and Pengjie Ren and Zhumin Chen and Dawei Yin and Zhaochun Ren},
      year={2024},
      eprint={2304.09542},
      archivePrefix={arXiv},
      primaryClass={cs.CL},
      url={https://arxiv.org/abs/2304.09542}, 
}

@misc{deepseekai2026deepseekv4,
      title={DeepSeek-V4: Towards Highly Efficient Million-Token Context Intelligence},
      author={DeepSeek-AI},
      year={2026},
}

@misc{deepseekai2024deepseekv32,
      title={DeepSeek-V3.2-Exp: Boosting Long-Context Efficiency with DeepSeek Sparse Attention}, 
      author={DeepSeek-AI},
      year={2025},
}

@article{zhi2026compress,
  title={Compress-then-Rank: Faster and Better Listwise Reranking with Large Language Models via Ranking-Aware Passage Compression},
  author={Zhi, Zhewei and Zhang, Yingyi and Jing, Yizhen and Li, Xianneng and Liu, Jianing and Liu, Huajie and Ding, Yongliang},
  year={2026}
}

@inproceedings{liu2025leveraging,
  title={Leveraging passage embeddings for efficient listwise reranking with large language models},
  author={Liu, Qi and Wang, Bo and Wang, Nan and Mao, Jiaxin},
  booktitle={Proceedings of the ACM on Web Conference 2025},
  pages={4274--4283},
  year={2025}
}

@inproceedings{first,
  title={FIRST: Faster improved listwise reranking with single token decoding},
  author={Reddy, Revanth Gangi and Doo, JaeHyeok and Xu, Yifei and Sultan, Md Arafat and Swain, Deevya and Sil, Avirup and Ji, Heng},
  booktitle={Proceedings of the 2024 Conference on Empirical Methods in Natural Language Processing},
  pages={8642--8652},
  year={2024}
}
